\documentclass[journal]{IEEEtran}
\ifCLASSINFOpdf
\else
\fi
\usepackage{cite}
\usepackage{graphicx}
\usepackage{amsmath}
\usepackage{amsfonts,amssymb}
\usepackage{mathrsfs}
\usepackage{array}
\usepackage{multirow}
\usepackage{longtable}
\usepackage{rotating}
\usepackage{float}
\usepackage{booktabs}

\begin{document}
\title{Transformer-based Spatial-Temporal Feature Learning for EEG Decoding}

\author{Yonghao Song, Xueyu Jia, Lie Yang,
        and Longhan~Xie,~\IEEEmembership{Member,~IEEE} 
\thanks{(corresponding author: Longhan Xie)}
\thanks{Yonghao Song, Xueyu Jia, Lie Yang and Longhan Xie are with the Shien-Ming Wu School of Intelligent Engineering, South China University of Technology, Guangzhou 510460, China (e-mail: eeyhsong@gmail.com, xielonghan@gmail.com). }}

\markboth{}%
{Shell \MakeLowercase{\textit{et al.}}: Bare Demo of IEEEtran.cls for IEEE Journals}

\maketitle

\begin{abstract}
At present, people usually use some methods based on convolutional neural networks (CNNs) for Electroencephalograph (EEG) decoding. However, CNNs have limitations in perceiving global dependencies, which is not adequate for common EEG paradigms with a strong overall relationship. Regarding this issue, we propose a novel EEG decoding method that mainly relies on the attention mechanism. The EEG data is firstly preprocessed and spatially filtered. And then, we apply attention transforming on the feature-channel dimension so that the model can enhance more relevant spatial features. The most crucial step is to slice the data in the time dimension for attention transforming, and finally obtain a highly distinguishable representation. At this time, global averaging pooling and a simple fully-connected layer are used to classify different categories of EEG data. Experiments on two public datasets indicate that the strategy of attention transforming effectively utilizes spatial and temporal features. And we have reached the level of the state-of-the-art in multi-classification of EEG, with fewer parameters. As far as we know, it is the first time that a detailed and complete method based on the transformer idea has been proposed in this field. It has good potential to promote the practicality of brain-computer interface (BCI).  
The source code can be found at: \textit{https://github.com/anranknight/EEG-Transformer}.
\end{abstract}


\IEEEpeerreviewmaketitle

\section{Introduction}
%
%
%
%
\IEEEPARstart{B}{rain}-computer interface (BCI) is a technology that provides the possibility of establishing direct connections between the brain and external devices \cite{Intro_1}. Researchers extract brain signals to obtain the user's intention, which is further used to control some equipment such as wheelchairs, robots and automatic vehicles \cite{Intro_2, Intro_3, Intro_4}. With BCI, it is easier to perform many tasks well without too much manpower. Especially it could help the disabled or paralyzed patients to get rid of the predicament that they have to rely on caregivers all the day \cite{Intro_5}.

For the above applications, one of the most widely studied BCI patterns are the sensorimotor rhythm stimulated by motor imagery (MI) on the motor cortex \cite{Intro_6}. Different categories of MI, such as right-hand movement and left-hand movement could be decoded with Electroencephalograph (EEG), an economical and convenient method for measuring information in the brain \cite{Intro_7}. Then the results are used as commands to control assistive devices and support the user to do corresponding movement. Rehabilitation training based on MI-BCI has been proven to be effective in helping patients suffering from strokes recover \cite{Intro_8, Intro_9}. Obviously, these BCI scenarios and applicable groups all require sufficient stability and reliability, which means that the decoding of brain signals should be accurate and robust.

Researchers have tried a lot of decoding methods to classify different EEG signals. The two main concerns are classification and feature extraction. In the beginning, some traditional machine learning methods such as linear discriminant analysis (LDA), support vector machine (SVM), are applied to analyze feature distribution and find a projection or hyper-plane to separate different categories \cite{Intro_10, Intro_11}. From another perspective, multi-layer perceptron (MLP) is also used to fit the input-output relationship with several hidden layers. The input is EEG signal, and the output is the category label \cite{Intro_12}. Although these methods are quite efficient, with the advancement of acquisition equipment, the large-scale data we obtain is difficult to handle properly. The ensuring amounts of irrelevant noise also challenge their generalization ability \cite{Intro_13}. 

Subsequently, deep learning methods have achieved remarkable results in computer vision and neural language processing. Features in multiple small fields are well perceived by the convolutional neural networks (CNNs) to obtain deep representation for classification. People also use CNNs in BCI to establish end-to-end EEG decoding models and achieve leading performance \cite{Intro_14, Intro_15}. Some skillful network structures use different scales of convolution calculations to extract features in various domains \cite{Intro_16, Intro_16.5}. However, CNN is very dependent on the selection of kernels. The large kernel hinders its exploration of deep feature, while the small kernel limits the receptive field \cite{Intro_17}. It loses part of time-series information and is difficult to perceive a wide range of internal relationships of the signal without a fairly deep structure, which may cause a lot of computations. Various recurrent neural networks (RNNs) are proposed to learn the temporal feature of EEG signal and have gained advantages in some scenarios \cite{Intro_18}. For instance, long short-term memory networks (LSTMs) employ gates for controlling the state of information flow to learn long-term dependencies of EEG signal. Nevertheless, these methods are still not enough to deal with more extended data, and less efficient due to RNN steps cannot be parallelized \cite{Intro_19}. Another problem is RNNs only act on the previous memory and current state. But we know that the action of a trial of EEG is coherent. In other words, every part of the action is related, including past and present, present and future \cite{Intro_20}. In this case, neither CNNs nor RNNs are enough to perceive global dependencies of EEG well. Recently, there is a method called self-attention applied in machine translation, which calculates the representation of a sequence with dependencies between different positions \cite{Intro_21}. It seems that the attention mechanism might help us decode EEG signals more reasonably. 

Another concern that cannot be ignored is feature extraction. EEG is a non-invasive brain data acquisition mode, which introduces a lot of noise caused by large impedance, field potential interference, and so on \cite{Intro_22}. Even those excellent end-to-end classification methods are difficult to obtain good performance without well-designed modules for feature extraction directly. Therefore, people are also trying to find more distinguishing features to facilitate EEG decoding. Fast Fourier transform (FFT) is used to convert the original EEG signal to a frequency representation, and continuous wavelet transform (CWT) is used for time-frequency features \cite{Intro_23, Intro_24}. Among these strategies, common spatial pattern (CSP), which focus on spatial features, is the most widely recognized \cite{Intro_25}. In this way, the original signal is filtered to a new space with larger differences between two categories. Earlier, small-scale feature vectors are calculated for classification, and later the filtered signal with temporal feature retained is employed as the input of CNN and other networks \cite{Intro_26}. But there is often a shortcoming that for multi-classification tasks, these methods simply stack feature channels obtained by multiple results of one-versus-rest (OVR) processing, and then use a convolution kernel with a size corresponding to the feature channels to handle these spatial features. This integration strategy largely neglects the importance of different feature channels, so collaborative optimization is not well performed. Especially for some data with minor category differences, it may causes serious mutual interference.

To address the above problems, in this paper, we apply the attention mechanism to construct a tiny transformer for EEG decoding, which means mainly depends on attention to transform the input into a more distinguishable representation. This model is called Spatial-Temporal Tiny Transformer (S3T) because it focuses on capturing spatial and temporal features. Firstly, the original EEG signal is filtered following the idea of CSP, and the outputs of multiple OVRs are stacked to form the input of S3T. Secondly, a feature-channel attention block gives weights to different channels so that the model could selectively pay attention to more relevant channels and ignore others. After that, all channels are compressed to reduce computation cost. Thirdly, the data is divided into many small slices in the time dimension. The attention mechanism is used to obtain a representation suitable for classification, by perceiving global temporal features. Finally, we get the category label with a simple fully-connected layer after global average pooling. Besides, we innovatively use a convolution-based layer to retain the position information of the sample. During training, only cross-entropy is employed as the constrain for optimization.

The major contributions of this article can be summarized as follows.

1) We propose a tiny framework named S3T for EEG decoding that mainly relies on the attention mechanism to learn the spatial and temporal features of EEG signals. It has the potential as a new backbone to classify EEG, like CNNs.

2) We present a strategy to weight feature channels, thereby improving the limitation of previous methods that neglect the importance of different feature channels.

3) Detailed experiments on public datasets prove the competitiveness of our method, reaching the level of the state-of-the-art with fewer parameters.

The remainder of this paper is organized as follows. Some related works are given in Section II. Details of our framework S3T is explained in Section III. We introduce the datasets, experiment settings, performance evaluation, comparative results in Section IV. A careful discussion is in Section V. Finally, we come to a conclusion in Section VI.

\section{Related works}
The development of EEG decoding methods could be summarized in two stages, the first of which is to find some representative patterns and use classical machine learning for classification. We will not go into details since the previous articles already have pretty good descriptions \cite{Related_1,Related_2}. And it comes to the second stage when the development of hardware promotes deep learning. The CNN-based models have a great ability to perceive feature dependence within a specific range and deepen it layer by layer. Some methods have obtained good EEG decoding results with the help of CNNs. Sakhavi et al. spatially filtered EEG signal and sampled it in time to get the input of a CNN, where two-scale kernels were used to capture information in different domains \cite{Related_3}. Zhao et al. transformed the data into a 3-dimensional (3D) form as the spatial distribution of the channels to cooperate with a 3D CNN to better use the original characteristics \cite{Related_4}. Some special objective functions are also used to improve the difference of different categories \cite{Related_5}. We can see that different kernels have an essential impact on the learning of the features. And a kernel with a limited length is not easy to measure the relationship between long-distance features, even if there are many layers. Moreover, the amount of parameters to be calculated is usually considerable. On the other hand, some RNN-based methods also yield good results leveraging time-series information. Wang et al. used it to deal with a one dimension-aggregate approximation representation of the EEG signal for MI classification \cite{Related_6}. Ma et al. employed LSTM to learn the spatial features and temporal features separately \cite{Related_7}. Zhang et al. tried to concatenate CNN and LSTM to fuse the advantage of the two \cite{Related_8}. 

In recent year, new methods different from CNNs and RNNs called attention mechanism has appeared in some fields. Vaswani et al. creatively constructed an encoder and decoder mainly composed of attention modules to convert a sequence of sentences in one language to another \cite{Intro_21}. This strategy contains this idea shows a great potential to dig out more expressive global dependencies. Dosovitskiy et al. divided the picture into multiple patches and used attention modules to obtain a token related to category information \cite{Related_9}. Some people have also partly introduced the attention mechanism to EEG decoding. Zheng et al. replaced the forget gate of LSTM and weighted the output with attention modules \cite{Related_10}. Tao et al. integrated attention with CNN and RNN to explore more discriminative information of EEG for emotion recognition \cite{Related_11}. Zhang et al. respectively processed the sliced EEG signal and a graph with CNN and LSTM then utilized attention module to synthesize all slices for classification \cite{Related_12, Related_13}. These remarkable methods have demonstrated some capability of the attention mechanism. Therefore, we try to further explore its ability to perceive features and use it as a trunk to construct S3T.

\section{Methods}
\begin{figure*}[htpb]
\centering
\includegraphics[width=\linewidth]{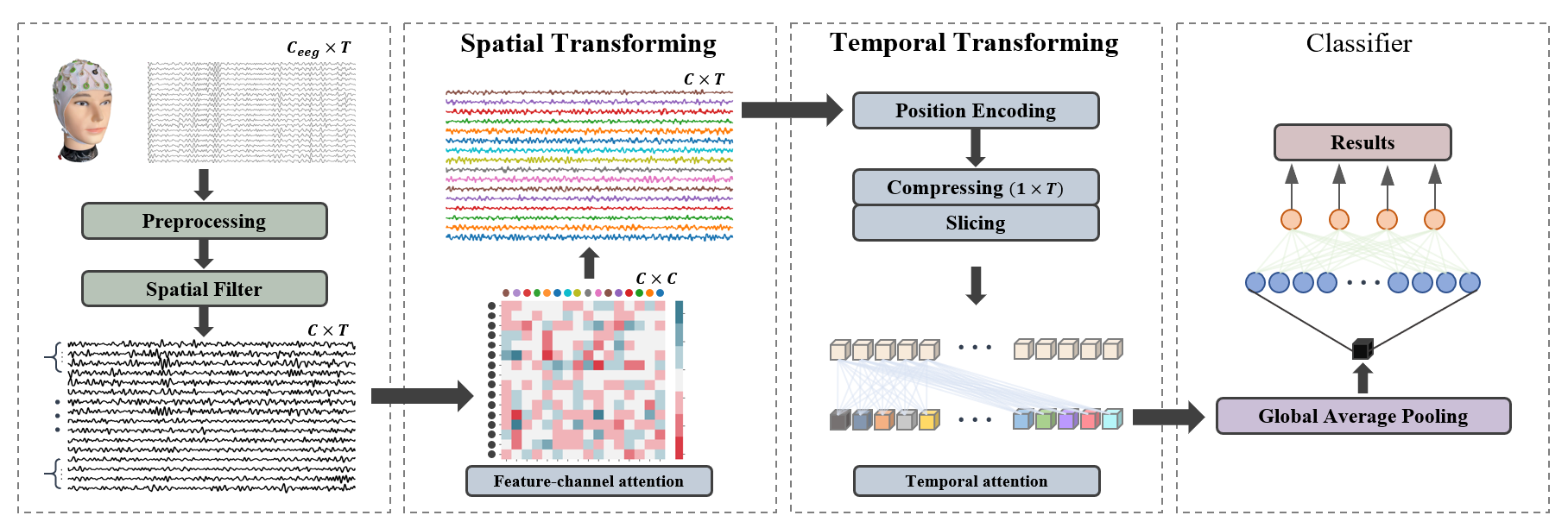}
\caption{The overall framework of Spatial-Temporal Tiny Transformer (S3T). After preprocessing and spatial filter, the method contains spatial transforming and temporal transforming built with the attention mechanism. There is a simple classifier consisting of global average pooling and a fully-connected layer.}
\label{fig:1}
\end{figure*}

In this paper, we propose an EEG decoding method named Spatial-Temporal Tiny Transformer (S3T). Just like a Rubik's Cube, S3T relies on the attention mechanism to transform the data into a highly distinguishing representation with the help of spatial and temporal information. The global dependencies of the signal and the importance of different feature channels are well perceived for classification tasks. The cost effectiveness of this method is also worth noting. 

The overall framework is shown in Fig. \ref{fig:1}, including four parts. The original EEG signal is firstly preprocessed, and the spatial filter is calculated to improve the difference between different categories. Then in the spatial transforming part, we apply the feature-channel attention to weight the feature-channels so that the model is able to focus on more relevant channels and ignore irrelevant ones. Thirdly, in the temporal transforming part, we use a layer of convolution to encode the position information in the time domain. After that, the data is compressed in a single channel to reduce computation cost. Meanwhile, it is divided into multiple small slices. The same size representation is inputted to the classifier after learning the global dependencies with temporal attention. Finally, all the slices are merged by global average pooling, and the result is obtained through a simple fully connected structure. Next, we will present the details of the four parts in S3T.   

\subsection{Preprocessing}
The preprocessing of raw EEG data includes segmenting, band-pass filtering, and standardization. Some approaches of calibration and artifact removal have been omitted to reduce excessive manual work.

The collected data is segmented into trials according to a whole MI process. The shape of a trial is expressed as $C_{eeg}\times T$, where $C_{eeg}$ is the number of EEG channels and $T$ is the sample points. Then we band-pass filter the data to [4, 40] Hz to remove high and low-frequency noise, during which the sensorimotor rhythms is disentangled, retaining $\mu$ and $\beta$ band \cite{Method_1}. The z-score standardization is employed to relieve the fluctuation and nonstationarity as
\begin{equation}
    X=\frac{x-\mu}{\sqrt{\sigma^2}}
\end{equation}
where $X \in \mathbb R^{C_{eeg}\times T}$ and $x \in \mathbb R^{C_{eeg}\times T}$ represent the standardized and input signal. $\mu$ and $\sigma^2$ denote the mean value and variance of training data. After standardization, signals become normally distributed with a mean of 0 and a standard deviation of 1. The mean and variance will be used directly in the test.

\subsection{Spatial Filter}
In this section, we give a feasible way based on CSP usage \cite{Intro_25} to improve the spatial difference of the original signal and maintain the temporal information. Besides, we introduce a one-versus-rest (OVR) strategy to handle multi-classification tasks, since the traditional CSP is suitable for only two categories. OVR refers to dividing a multi-classification into $N$ bi-classification tasks of one category and the remaining category, where $N$ equals the number of categories. The sub-filters gained by each OVR are combined to form the spatial filter at last.  

In one OVR, we calculate the covariance matrix of each trial as $C(X)$, where $C()$ denotes calculate the covariance matrix. $X \in \mathbb R^{C_{eeg}\times T}$ is the preprocessed EEG data.

After that, we average all the covariance matrix of one category to obtain $R_1 \in \mathbb R^{C_{eeg}\times C_{eeg}}$ of the 'one' and $R_2 \in \mathbb R^{C_{eeg}\times C_{eeg}}$ of the 'rest'. In this way, spatial characteristics of the two categories are depicted because each position of the covariance matrix measures the jointly variability of two channels. Afterward, a common space $R \in \mathbb R^{C_{eeg}\times C_{eeg}}$ is obtained by
\begin{equation}
    R = R_1 + R_2
\end{equation}
And eigendecomposition is performed as 
\begin{equation}
    R = U\Lambda U^T
\end{equation}
where $U$ and $\Lambda$ denote the eigenvectors and the eigenvalues, respectively. $\Lambda$ is arranged in descending order. Then the whitening matrix $P$ of $R$ is obtained: 
\begin{equation}
    P = \sqrt{\Lambda^{-1}} U^T
\end{equation}
Applying the matrix $P$ to $R_1$ and $R_2$, we get
\begin{equation}
    \begin{aligned}
        & S_1 = P R_1 P^T \\
        & S_2 = P R_2 P_T \\
    \end{aligned}
\end{equation}
The orthogonal diagonalization of $S_2$ is obtained as
\begin{equation}
    S_2 = B \Lambda_S B^T
\end{equation}
where $B$ and $\Lambda_S$ are the eigenvectors and eigenvalues of $S_1$. $\Lambda_S$ is arranged in ascending order. Because of the orthogonality, there is 
\begin{equation}
    B^T B = I
\end{equation}
\begin{equation}
    \begin{aligned}
        I & = B^T \sqrt{\Lambda^{-1}} U^T U\Lambda U^T (\sqrt{\Lambda^{-1}} U^T)^T B \\
        & = B^T P R P^T B \\
        & = B^T P R_1 P^T B + B^T P R_2 P^T B \\
        & = B^T P R_1 P^T B + \Lambda_S \\
    \end{aligned}
\end{equation}
where $B^T P R_1 P^T B$ is a diagonal matrix, denoted as $\Lambda'_S$ with a value of $I-\Lambda_S$, which means that $B^T P$ diagonalizes both $R_1$ and $R_2$. So the values of $\Lambda'_S$ become larger when the values of $\Lambda_S$ becomes smaller. $B^T P$ is treated as a spatial filter, with which the difference between the 'one' and the 'rest' is maximized. 

We obtain $N$ outputs here, and the first $S$ rows of each are taken out as a sub-filter corresponding to the largest four eigenvalues of $\Lambda'_S$, to reduce the computational complexity. Then the data is filtered as
\begin{equation}
    Z = W X
\end{equation}
where $Z$ is the filtered data and $X$ is the preprocessed data. $W \in \mathbb R^{NS \times C_{eeg}}$ is the final spatial filter made by stacking $N$ sub-filters.

\subsection{Spatial Transforming}
\begin{figure}[htpb]
    \centering
    \includegraphics[width=0.5\linewidth]{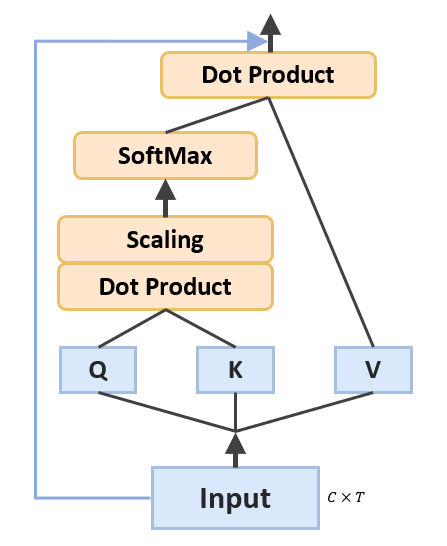}
    \caption{The calculation process of spatial feature-channel attention.}
    \label{fig:2}
\end{figure}

In previous methods, people rarely paid attention to the importance of different feature channels, leading to inefficiency and mutual interference of features. Therefore, we present a method of feature channel weighting inspired by scaled dot-product attention \cite{Intro_21}. The dependence of each element on other elements is learned to obtain the importance score for weighting the values of data. Specifically, we use dot product to evaluate the correlation between one feature channel and others as Fig. \ref{fig:2}. The input data is first linearly transformed into vectors, queries ($Q$) and keys ($K$) of dimension $d_k$, and values ($V$) of dimension $d_v$, along the spatial feature dimension. $Q$ represents each channel that will be used to match with $K$ represents all the other channels using dot product. Then the result is divided by a scaling factor of $\sqrt{d_k}$ to ensure that the $Softmax$ function has a good perception ability. The output weight score is assigned to $V$ for the final representation using dot product. The whole process can be expressed as
\begin{equation}
    Attention(Q, K, V) = Softmax(\frac{QK^T}{d_k})V
\end{equation}
where $Attention(Q,K,V)$ is the weighted representation. $Q$, $K$ and $V$ are matrices packed by vectors for simultaneous calculation. Besides, the residual connection of the input and the output is used to help gradients of the framework flow \cite{Method_2}.

\subsection{Temporal Transforming}
Here is the most important part of this article, that is, perceiving global temporal dependencies of EEG signals using the attention mechanism. We know that a behavior driven by the brain is a complete process, so our method could be more effective to utilize the relationship between any two parts of a trial. In the beginning, the data is compressed to one dimension ($1 \times T$) to reduce the computational complexity, since the feature channels have been weighted in the previous step. Through the observation of EEG signal, a segment reflects the trend better than a single sample point, which is more likely to have abnormal values. Therefore, we divide the data into multiple slices with a shape of $1 \times d$ for attention training. Different from spatial transforming, multi-head attention (MHA) \cite{Intro_21} is employed to allow the model to learn the dependencies from different angles. In this way, the input is split into $h$ smaller parts named heads, which will perform attention in parallel. And the outputs of each part are concatenated and linearly transformed to obtain the original size. The process can be expressed as 
\begin{equation}
    {MHA}(X_Q, X_K, X_V) = [head_0;...;head_{h-1}]W^o
\end{equation}  
\begin{equation}
    head_i=Attention(X_Q W^Q_i, X_K W^K_i, X_V W^V_i)
\end{equation}  
where $[.;.]$ represents a concatenation operation. $W^Q_i \in \mathbb R^{d \times \frac{d_k}{h}}$, $W^K_i \in \mathbb R^{d \times \frac{d_k}{h}}$ and $W^V_i \in \mathbb R^{d \times \frac{d_v}{h}}$ denote linear transformations to obtain queries, keys and values matrices of each head. $ W^o \in \mathbb R^{d_v \times d }$ denotes the linear transformation to obtain the final output. A feed-forward (FF) block contains two fully-connected layers and the activation function GeLU \cite{Mehtod_3} is connected behind the MHA, to enhance the perception and non-linear learning capabilities of the model. The input and output sizes of the FF block are the same, and the inner size is expanded to $N_f$ times. Layer normalization \cite{Method_4} is set before MHA and FF block. The residual connection is also used for better training. The module with MHA and FF block is repeated $N_a=3$ times for a ensemble effect.  
Although the temporal transforming measure the dependencies between different slices well, it ignores the position information, which is the sequence relationship between the EEG sample points. We creatively use a convolutional layer on time dimension with a kernel size of $k_c$ and stride of $1$ to encode position information before compressing and slicing.  

\subsection{Classifier}
After the above process, the data is transformed into a new representation by learning the spatial and temporal dependencies. Now we just need to apply a global pooling to average all the slices in the temporal transforming part. And the pooling result is connected to a fully-connected layer after layer normalization. The number of output neurons is equal to the number of categories. Then, the Softmax function is used to obtain the predicted probability. The objective function is the classification loss achieved by cross-entropy as 
\begin{equation}
    \mathcal{L} = -\frac{1}{M}\sum^{N}_{n=1}\sum^{M}_{m=1}y^n_m log(\hat{y_m}^n) 
\end{equation} 
where $M$ is the number of trials and $N$ is the number of categories. $y^m_n$ denotes the real label for the m-th trial, and $\hat{y_m}^n$ represents the predicted probability of the m-th trial for the category n.

\section{Experiments and Results}
\subsection{Datasets}
To evaluate the proposed S3T model, we conducted detailed experiments on two public datasets, datasets 2a and 2b of BCI competition IV.
\subsubsection{Datasets 2a of BCI Competition IV}
The datasets \cite{Exp_1} provided by Graz University of Technology were used for our experimental study. The MI EEG data of nine subjects were acquired with twenty-two Ag/AgCl electrodes at a sampling rate of 250 Hz, and band-pass filtered between 0.5 Hz and 100 Hz. There were four different tasks, including imagination of moving left hand, right hand, both feet, and tongue. Two sessions of data were collected for each subject, and each session contained 288 trials (72 for each task). The temporal segment of [2, 6] second was used in our experiments.
\subsubsection{Datasets 2b of BCI Competition IV}
The datasets \cite{Exp_2} recorded the MI data of nine subjects on three bipolar electrodes. The data were band-pass filtered between 0.5 Hz and 100 Hz with a sampling rate of 250 Hz. The imagination of moving left hand and right hand was applied to be the paradigm. Two sessions without visual feedback and three sessions with visual feedback were obtained for every subject. We used 120 trials in each session, due to the number of trials for several sessions was slightly different. The segment of [3, 7] second was maintained as one trial in the experiments.

\subsection{Experiment details}
Our method was implemented with Python 3.6 and PyTorch library on a Geforce 2080Ti GPU. The electrooculogram channels of the data were removed directly, and there was no additional artifact removal except for band-pass filtering. We trained subject-specific models according to the usual implementation. Ten-fold cross-validation was used to evaluate the final results. The slice size, $h$, $k_c$, and $N_f$ were 10, 5, 51 and 4, respectively. For the first datasets, category number $N$ was 4, and the row number $S$ applied in the spatial filter was 4. In the second dataset, there was $N=2$ and just one sub-filter was obtained with $S=3$. In the model, Adam \cite{Exp_3}, with a learning rate of 0.0002, was utilized to optimize the training process. $\beta_1$, $\beta_2$, and batch size were 0.5, 0.9, 50 separately. We introduced a dropout of 0.5 in temporal transforming and 0.3 in spatial transforming to avoid overfitting. Wilcoxon Signed-Rank Test was used for statistical analysis.

\subsection{Scoring performance}
\begin{figure}[htpb]
    \centering
    \includegraphics[width=0.9\linewidth]{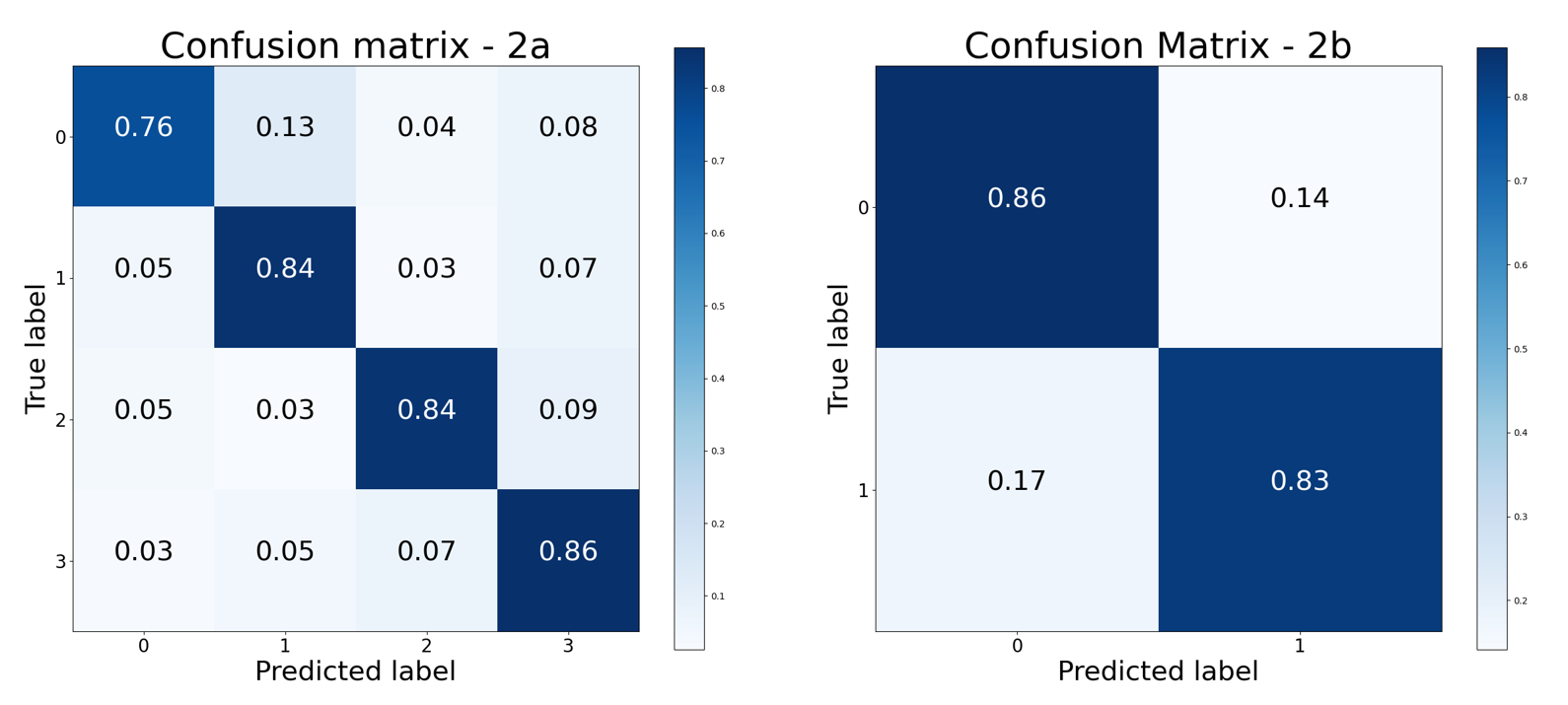}
    \caption{The classification results on BCI competition IV datasets 2a and BCI competition IV datasets 2b shown in confusion matrices.}
    \label{fig:3}
\end{figure}
\begin{table}[htbp]
\begin{center}
    \caption{Scoring performance of S3T.}   
    \begin{tabular}{clccccc}    
        \toprule    
             & label & Accuracy & Precision & Recall & Specificity & F-score \\    
        \midrule 
        \multirow{4}*{2a}
        & c0 & 91.30 & 83.33 & 75.60 & 95.72 & 79.28  \\   
        & c1 & 91.48 & 81.88 & 84.33 & 93.84 & 83.09  \\   
        & c2 & 92.03 & 87.84 & 83.87 & 95.32 & 85.81  \\   
        & c3 & 90.37 & 77.40 & 85.61 & 91.91 & 81.30  \\ \midrule  
        \multirow{2}*{2b}
        & c0 & 84.26 & 83.09 & 85.87 & 82.66 & 84.46  \\   
        & c1 & 84.26 & 85.50 & 82.66 & 85.87 & 84.06  \\   
        \bottomrule   
    \end{tabular} 
    \label{tab:1}
\end{center}
\end{table}
In the beginning, we give an overview with several metrics to show the performance of S3T. Confusion matrices of the classification results on two datasets are displayed in Fig. \ref{fig:3}, where the horizontal axis represents the predicted label and the vertical axis represents the true label. We further evaluate the method with five indicators for each category, accuracy, precision, recall, specificity, and F-score, as given in Table \ref{tab:1}. These matrices are calculated in OVR mode, considering one category as the positive and the other as the negative. F-score could be calculated as
\begin{equation}
    F_{score} = (2 \times Recall) / (Precision + Recall)
\end{equation}
It can be seen that the overall performance of our method is trustworthy, with good classification ability for different categories and not much bias.   

\subsection{Statistical Analysis with Baselines}
\begin{table*}[htbp]
\begin{center}
	\caption{Comparison with representative methods on datasets 2a and 2b of BCI competition IV.}
	\setlength{\tabcolsep}{4pt} 
	\renewcommand\arraystretch{1.2} 
	\begin{tabular}{ccccccccccccccc}
		\toprule   
	    datasets & methods & S01 & S02 & S03 & S04 & S05 & S06 & S07 & S08 & S09 & Average & std & significance & Params \\ 
		\midrule
		\multirow{7}*{2a}
		& FBCSP \cite{Exp_4} & 76.00 & 56.50 & 81.25 & 61.00 & 55.00 & 45.25 & 82.75 & 81.25 & 70.75 & 67.75 & 12.94 & \textit{p}\ \textless\ 0.01 & --- \\
		& ConvNet \cite{Exp_5} & 76.39 & 55.21 & 89.24 & 74.65 & 56.94 & 54.17 & 92.71 & 77.08 & 76.39 & 72.53 & 13.42 & \textit{p}\ \textless\ 0.01 & 295.25k \\
		& EEGNet \cite{Exp_6} & 85.76 & 61.46 & 88.54 & 67.01 & 55.90 & 52.08 & 89.58 & 83.33 & 86.81 & 74.50 & 14.36 & \textit{p}\ \textless\ 0.01 & 1.46k \\ 
		& C2CM \cite{Related_3} & 87.50 & 65.28 & 90.28 & 66.67 & 62.5 & 45.49 & 89.58 & 83.33 & 79.51 & 74.46 & 14.45 & \textit{p}\ \textless\ 0.01 & 36.68k \\
		& CNN+LSTM \cite{Exp_7} & 85.00 & 54.00 & 87.00 & 78.00 & 77.00 & 66.00 & 95.00 & 83.00 & \textbf{90.00} & 80.00 & 11.97 & \textit{p}\ = 0.0961 & 8.57k \\
		& DFL \cite{Related_5} & 91.31 & 71.62 & 92.32 & \textbf{78.38} & \textbf{80.10} & 61.62 & 92.63 & 90.30 & 78.38 & 81.85 & \textbf{10.15} & \textit{p}\ = 0.0774 & 30.69k \\ 
		& \textbf{Ours} & \textbf{91.67} & \textbf{71.67} & \textbf{95.00} & 78.33 & 61.67 & \textbf{66.67} & \textbf{96.67} & \textbf{93.33} & 88.33 & \textbf{82.59} & 12.52 & --- & 8.68k \\ \midrule
		\multirow{5}*{2b}
		& FBCSP \cite{Exp_4} & 70.00 & 60.36 & 60.94 & 97.50 & 93.12 & 80.63 & 78.13 & 92.50 & \textbf{86.88} & 80.00 & 13.06 & \textit{p}\ \textless\ 0.05 & --- \\
		& ConvNet \cite{Exp_5} & 76.56 & 50.00 & 51.56 & 96.88 & \textbf{93.13} & 85.31 & 83.75 & 91.56 & 85.62 & 79.37 & 16.27 & \textit{p}\ \textless\ 0.05 & 295.23k \\
		& EEGNet \cite{Exp_6} & 68.44 & 57.86 & 61.25 & 90.63 & 80.94 & 63.13 & 84.38 & 93.13 & 83.13 & 75.88 & 12.57 & \textit{p}\ \textless\ 0.01 & 1.15k \\ 
		& MSCNN \cite{Exp_8}  & 80.56 & 65.44 & 65.97 & \textbf{99.32} & 89.19 & 86.11 & 81.25 & 88.82 & 86.81 & 82.61 & 10.44 & \textit{p}\ \textless\ 0.05 & 24.99k\\
		& \textbf{Ours} & \textbf{81.67} & \textbf{68.33} & \textbf{66.67} & 98.33 & 88.33 & \textbf{90.00} & \textbf{85.00} & \textbf{93.33} & 86.67 & \textbf{84.26} & \textbf{10.03} & -- & \textbf{6.50k}  \\
		\bottomrule
	\end{tabular}
	\label{tab:2}
\end{center}
\end{table*}

Then we conduct comparative experiments and perform significance analysis with some recent representative baselines. Brief descriptions are given as follows.

\begin{itemize}
\item \textit{FBCSP} \cite{Exp_4} is one of the most widely accepted methods that applies the filter bank strategy to cooperate with CSP. 
\item \textit{ConvNet} \cite{Exp_5} presents a remarkable demonstration to use CNN-based deep learning models for EEG decoding.
\item \textit{EEGNet} \cite{Exp_6} designs a compact and practical CNN with depthwise and separable convolutions  to classify EEG.
\item \textit{C2CM} \cite{Related_3} introduces a temporal representation with CSP and utilize a CNN architecture for classification.
\item \textit{CNN+LSTM} \cite{Exp_7} segments the original signal for OVR-CSP processing and extracts features with CNN, and the output is passed through an LSTM for temporal features.
\item \textit{DFL} \cite{Related_5} uses a well-designed feature separation strategy in CNN to improve EEG decoding performance.
\end{itemize} 
The accuracy of each subject, average accuracy, standard deviation (std), parameter amount (Params) on two datasets are listed in Table \ref{tab:2}. P-values are used to measure the significance of our method. We can see that the accuracy of our S3T relies on attention mechanism has obvious superiority with a limited amount of parameters on the two datasets, and the std is at a low level, which means that the method is effective and robust for EEG decoding.

To be specific, in datasets 2a, FBCSP makes good use of CSP, but the accuracy is lower because it compresses the time domain and uses traditional machine learning classifiers. ConvNet and EEGNet show the power of CNN-based deep networks in EEG classification. Although the convolutions of different scales are used dexterously to extract spatial domain and temporal domain features, the performance is still inefficient due to the lack of perception of global dependencies. C2CM combines the advantages of CSP and CNNs but has similar limitations. CNN+LSTM further introduces LSTM to deal with temporal dependencies, and DFL employs additional objective functions to expand feature differences. However, except for the weak results on Subject 05, our method still has a significant improvement than CNN+LSTM (\textit{p}\ = 0.0210) and DFL (\textit{p}\ = 0.0104). Another important issue is that the methods that rely on CNNs usually have a large number of parameters, such as ConvNet, C2CM, and DFL. Fewer parameters of our method reduce the computational complexity and cost to a large extent. EEGNet has few parameters but insufficient performance. The comparison in datasets 2b also shows a consistent trend, and the std of our method is lower than all others. 

The above results illustrate that our method has quite great performance and is more cost-effective.

\subsection{Ablating Study}
\begin{figure}[htpb]
    \centering
    \includegraphics[width=\linewidth]{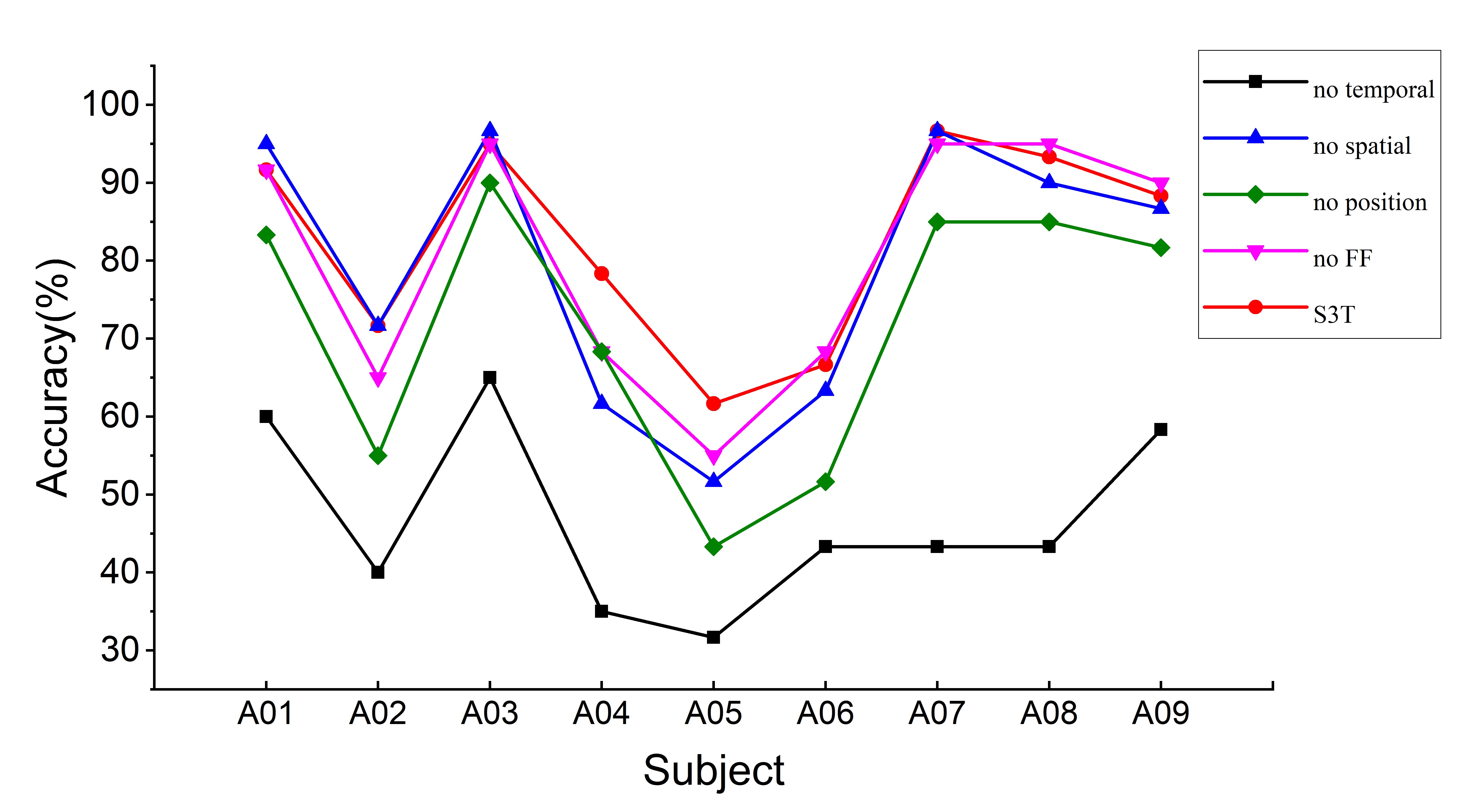}
    \caption{The results of the ablation study on datasets 2a.}
    \label{fig:4}
\end{figure}
The core idea of the proposed S3T method is to transform the EEG signal into a distinguishable representation by capturing spatial and temporal features. Therefore, we present an ablation study on BCI competition datasets 2a to show the effectiveness of each part in S3T as Fig. \ref{fig:4}. 

We first remove the temporal transforming and spatial transforming module from the complete S3T, respectively. It can be seen from Fig. \ref{fig:4} that the results are in line with our expectations. Temporal transforming is the most important in the model, and its absence makes the mean accuracy have a significant drop of 35.93\%. The contribution of spatial transforming has also been proven to be sufficient, with the mean accuracy dropping 3.33\% when we remove it. Besides, spatial transforming does have a distinct improvement for the data with poor discrimination, such as subject 5 and 6.

Besides, we employ a convolutional layer as a position encoding block and a feed forward (FF) block after each MHA in the temporal transforming module. So these two blocks are discarded separately to test their contribution to the model. As shown in Fig. \ref{fig:4}, position encoding largely compensates for the use of only attention mechanism with an improvement of 11.11\%. It also demonstrates that position information is valuable for EEG decoding. FF blocks also give the model a more robust learning capability like spatial transforming. And the mean accuracy increases by 2.22\% after adding FF blocks. The above experiment explains that the main parts in S3T have the evident effect, where the temporal transforming is the trunk.

\subsection{Parameter Sensitivity}
\begin{figure}[htpb]
    \centering
    \includegraphics[width=\linewidth]{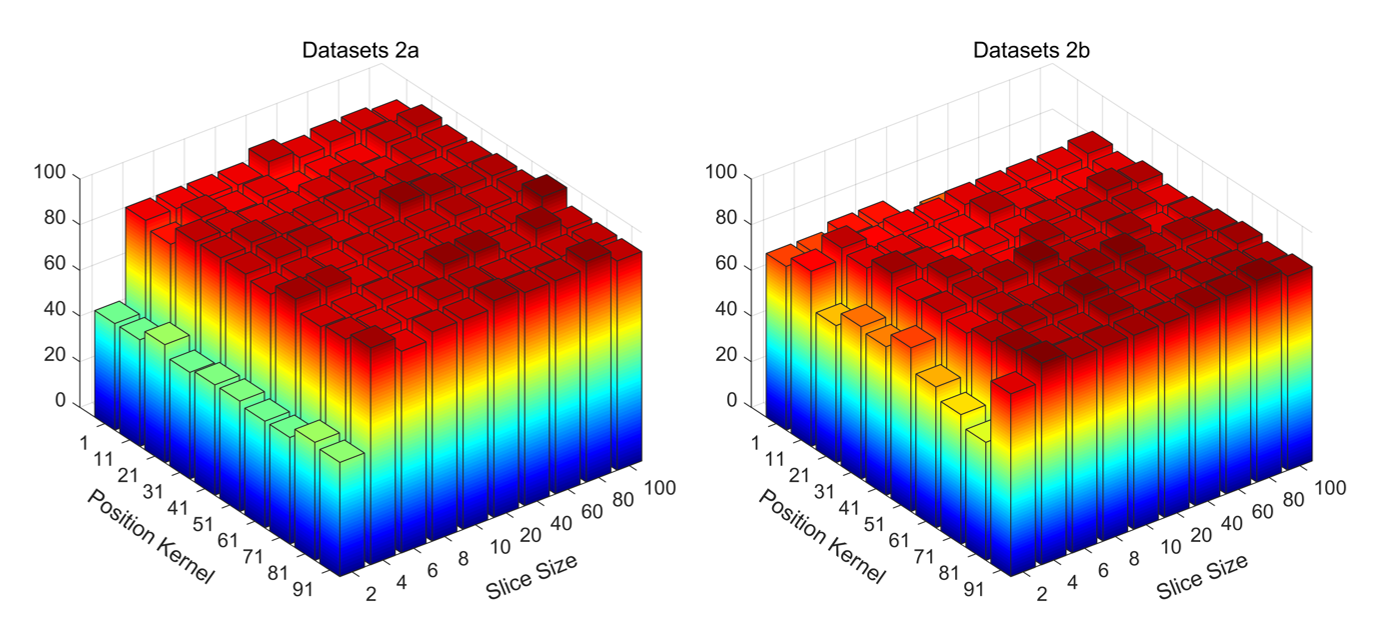}
    \caption{The results of parameter sensitivity tests. The position encoding kernel and the slice size are independently changed to test the performance of our model on the first subject of datasets 2a and datasets 2b.}
    \label{fig:5}
\end{figure}
It should be noted that two parameters may affect the performance of the model. They are the kernel size of position encoding and slice size in temporal transforming. We perform sensitivity analysis by varying these two parameters in a wide range and testing the impact on classification accuracy. The comparison results on the  data of the first subject in two datasets are depicted in Fig. \ref{fig:5}, separately. We can see that the results drop sharply when the slice size is small. In fact, we choose to slice, on the one hand, to reduce computation; another more important reason is to conduct attention on many short sequences that better reflect the state of the signal, just like CNNs do. Smaller slices are also more susceptible to some noise or outliers. 

For position encoding, the smaller kernel size reduces the performance expressly. We introduce position encoding to improve the perception of sequence position information, because the attention mechanism directly focuses on global dependencies. Therefore, when the kernel size is too small, it is hard to capture valuable information in a wide enough range. 

After the position kernel and slice size taking larger values, the result fluctuates slightly. So our method is confirmed to be stable enough when the parameters change.  

\subsection{Interpretability and Visualization}
\begin{figure*}[htpb]
    \centering
    \includegraphics[width=\linewidth]{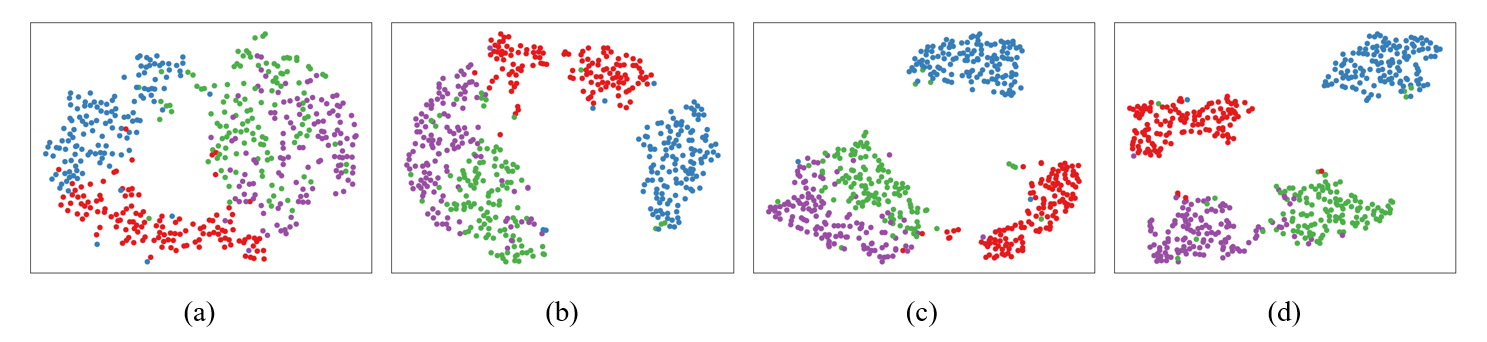}
    \caption{Visualization with t-SNE. (a) Data distribution after spatial transforming. (b) Data distribution after the first layer of temporal transforming. (c) Data distribution after the second layer of temporal transforming. (d) Data distribution after the third layer of temporal transforming, also the input of the classifier. }
    \label{fig:6}
\end{figure*}
The learning process is visualized with t-SNE \cite{Exp_9} to interpret the S3T model further. We first train the model to be stable and then visualize the data of one epoch in four stages, after spatial transforming and after each of the three temporal transforming layers as Fig. \ref{fig:6}. The data used to display is from subject 1 in BCI competition IV datasets 2a. The four colors in the figure represent four categories of EEG. Fig. \ref{fig:6}(a) is the output of the spatial transforming. We can see that the four categories have been mildly distinguished, but not very obvious, and there are two categories completely mixed together. According to Fig. \ref{fig:6}(b)-(d), after each layer of temporal transforming is processed, the distinction among the four categories gradually becomes larger, until these data can be clearly distinguished in the end.

\section{Discussion}
Better decoding is a necessary issue to improve the application of BCI. In recent years, CNN-based deep learning methods plays a dominant role in EEG analysis. But CNNs have a limitation on perceiving global dependencies of long sequences. Especially for EEG signal like motor or motor imagery, a trial is inseparable. Therefore, we propose a method that uses attention mechanism to deal with global dependencies of EEG signal. Several layers of attention operations transform the original input into a distinguishable representation for classification. Besides, we also perform attention on the feature channels of spatially filtered EEG, which helps us avoid aliasing of channels. After detailed experiments, we find that this transforming idea has the potential to compete with usual CNN and LSTM models in EEG decoding.

According to the ablation study in Fig. \ref{fig:4}, temporal transforming is the backbone of our method. We slice the data into many small segments for attention, which is inspired by the method in computer vision to divide the image into many patches \cite{Related_9}. Refer to Fig. \ref{fig:5}, this measure is more effective and with lower computation cost than directly processing sample points. The influence of some outliers is alleviated by observing a few continuous sample points. In this way, after several rounds of attention weighting, the model focuses on slices that are more related to category information with the constraints of only classification loss. So the final representation is easily separated as Fig. \ref{fig:6}(d). 

It should be noted that we don't completely abandon convolution as "attention is all you need" \cite{Intro_21}. Although the attention has well perceived the global dependencies of the EEG signal, the position encoding realized by a layer of convolution still plays an important role, as shown in Fig. \ref{fig:4}. The application of attention by slicing ignores the sequence relationship of the signal to a certain extent, and the convolution of a nearby area just makes up for this problem. Besides removing the convolution in ablation study, we can also see from Fig. \ref{fig:5} that too small convolution kernel also reduces the performance, proving that the position encoding implemented in this way is effective. New methods appear every day, and we are not going to dismiss the previous methods but try to combine the advantages of different strategies.

In temporal transforming, we compress the signal from the feature-channel dimension in 1 channel to reduce calculation. This is because the feature-channels have been weighted in spatial transforming with attention mechanism, inspired by \cite{Exp_10}. One explanation for its meaning is that OVR strategy and spatial filter are employed to process the original EEG signal like \cite{Related_3}. In this case, stacked feature-channels obtained from different subspaces may cause interference. So spatial transforming in advance could increase the model's attention to more relevant channels. Experiments display that it does enhance the overall performance, especially with a significant improvement on 'A05' and 'A06' in Fig. \ref{fig:4}, whose data are not easily distinguishable.

The scale of our model is an issue that cannot be ignored. Some methods have achieved impressive results with a large number of parameters, but require considerable computing power and training time. So we strive to control the number of parameters of S3T at a relatively low level, on the premise of good accuracy. This mainly due to two reasons. One is that we compress the feature-channels. An operation with the similar idea is also used after temporal transforming, which is to average all slices as pooling as \cite{Exp_11}, instead of using some fully-connected layers as usual. This trick could be used with a small-size fully-connected layer to classify, own to the good representation provided by spatial and temporal transforming.

Although S3T achieves very nice performance on EEG decoding tasks, the limitation of this work should also be considered. Only subject-specific experiments are conducted as the common way, and the cross-subject ability has not been well explored. Another limitation is that we choose the hyper-parameters of the model, such as dropout and batch size, just by some preliminary experiments, so the model may not reach the optimal state. In future work, we will verify the generalization ability of the method with cross-subject tests, and try to use some data augmentation strategies to improve the performance.

\section{Conclusion}
In this paper, we propose S3T, a tiny method relies on attention mechanism to perceive the spatial and temporal features of EEG signal for decoding tasks. Firstly, we use attention to weight the preprocessed and spatially filtered data along the feature-channel dimension, enhancing more relevant channels. It is also employed along the time dimension to perceive global dependencies. Experiments show that our method is robust enough to classify multiple categories of EEG, reaching the state-of-the-art level with much fewer parameters. The stability of the method and the effectiveness of each module have also been proven. We further utilize visualization tools to confirm that it is meaningful to transform data through spatial and temporal features learning. In summary, our method has good potential to complement commonly used deep learning methods for EEG decoding.

\section*{Acknowledgment}
This work was supported in part by the National Natural Science Foundation of China (Grant No. 52075177), Joint Fund of the Ministry of Education for Equipment Pre-Research (Grant No. 6141A02033124), Research Foundation of Guangdong Province (Grant No. 2019A050505001 and 2018KZDXM002), Guangzhou Research Foundation (Grant No. 202002030324 and 201903010028), Zhongshan Research Foundation (Grant No.2020B2020), and Shenzhen Institute of Artificial Intelligence and Robotics for Society (Grant No. AC01202005011).

\ifCLASSOPTIONcaptionsoff
  \newpage
\fi


\bibliographystyle{IEEEtran}
\bibliography{ref}

\begin{thebibliography}{10}
\providecommand{\url}[1]{#1}
\csname url@samestyle\endcsname
\providecommand{\newblock}{\relax}
\providecommand{\bibinfo}[2]{#2}
\providecommand{\BIBentrySTDinterwordspacing}{\spaceskip=0pt\relax}
\providecommand{\BIBentryALTinterwordstretchfactor}{4}
\providecommand{\BIBentryALTinterwordspacing}{\spaceskip=\fontdimen2\font plus
\BIBentryALTinterwordstretchfactor\fontdimen3\font minus
  \fontdimen4\font\relax}
\providecommand{\BIBforeignlanguage}[2]{{%
\expandafter\ifx\csname l@#1\endcsname\relax
\typeout{** WARNING: IEEEtran.bst: No hyphenation pattern has been}%
\typeout{** loaded for the language `#1'. Using the pattern for}%
\typeout{** the default language instead.}%
\else
\language=\csname l@#1\endcsname
\fi
#2}}
\providecommand{\BIBdecl}{\relax}
\BIBdecl

\bibitem{Intro_1}
F.~R. Willett, D.~T. Avansino, L.~R. Hochberg, J.~M. Henderson, and K.~V.
  Shenoy, ``High-performance brain-to-text communication via handwriting,''
  \emph{Nature}, vol. 593, no. 7858, pp. 249--254, 2021.

\bibitem{Intro_2}
A.~Cruz, G.~Pires, A.~Lopes, C.~Carona, and U.~J. Nunes, ``A {Self}-{Paced}
  {BCI} {With} a {Collaborative} {Controller} for {Highly} {Reliable}
  {Wheelchair} {Driving}: {Experimental} {Tests} {With} {Physically} {Disabled}
  {Individuals},'' \emph{IEEE Transactions on Human-Machine Systems}, vol.~51,
  no.~2, pp. 109--119, Apr. 2021.

\bibitem{Intro_3}
A.~Schwarz, M.~K. Höller, J.~Pereira, P.~Ofner, and G.~R. Müller-Putz,
  ``Decoding hand movements from human {EEG} to control a robotic arm in a
  simulation environment,'' \emph{Journal of Neural Engineering}, vol.~17,
  no.~3, p. 036010, May 2020.

\bibitem{Intro_4}
Y.~Song, W.~Wu, C.~Lin, G.~Lin, G.~Li, and L.~Xie, ``Assistive {Mobile} {Robot}
  with {Shared} {Control} of {Brain}-{Machine} {Interface} and {Computer}
  {Vision},'' in \emph{2020 {IEEE} 4th {Information} {Technology},
  {Networking}, {Electronic} and {Automation} {Control} {Conference}
  ({ITNEC})}.\hskip 1em plus 0.5em minus 0.4em\relax Chongqing, China: IEEE,
  Jun. 2020, pp. 405--409.

\bibitem{Intro_5}
E.~Tidoni, M.~Abu-Alqumsan, D.~Leonardis, C.~Kapeller, G.~Fusco, C.~Guger,
  C.~Hintermuller, A.~Peer, A.~Frisoli, F.~Tecchia, M.~Bergamasco, and S.~M.
  Aglioti, ``Local and {Remote} {Cooperation} {With} {Virtual} and {Robotic}
  {Agents}: {A} {P300} {BCI} {Study} in {Healthy} and {People} {Living} {With}
  {Spinal} {Cord} {Injury},'' \emph{IEEE Transactions on Neural Systems and
  Rehabilitation Engineering}, vol.~25, no.~9, pp. 1622--1632, Sep. 2017.

\bibitem{Intro_6}
G.~Pfurtscheller, C.~Brunner, A.~Schlögl, and F.~Lopes~da Silva,
  ``\BIBforeignlanguage{en}{Mu rhythm (de)synchronization and {EEG}
  single-trial classification of different motor imagery tasks},''
  \emph{\BIBforeignlanguage{en}{NeuroImage}}, vol.~31, no.~1, pp. 153--159, May
  2006.

\bibitem{Intro_7}
A.~Al-Saegh, S.~A. Dawwd, and J.~M. Abdul-Jabbar,
  ``\BIBforeignlanguage{en}{Deep learning for motor imagery {EEG}-based
  classification: {A} review},'' \emph{\BIBforeignlanguage{en}{Biomedical
  Signal Processing and Control}}, vol.~63, p. 102172, Jan. 2021.

\bibitem{Intro_8}
R.~Foong, K.~K. Ang, C.~Quek, C.~Guan, K.~S. Phua, C.~W.~K. Kuah, V.~A.
  Deshmukh, L.~H.~L. Yam, D.~K. Rajeswaran, N.~Tang, E.~Chew, and K.~S.~G.
  Chua, ``Assessment of the {Efficacy} of {EEG}-{Based} {MI}-{BCI} {With}
  {Visual} {Feedback} and {EEG} {Correlates} of {Mental} {Fatigue} for
  {Upper}-{Limb} {Stroke} {Rehabilitation},'' \emph{IEEE Transactions on
  Biomedical Engineering}, vol.~67, no.~3, pp. 786--795, Mar. 2020.

\bibitem{Intro_9}
R.~Mane, T.~Chouhan, and C.~Guan, ``{BCI} for stroke rehabilitation: motor and
  beyond,'' \emph{Journal of Neural Engineering}, vol.~17, no.~4, p. 041001,
  Aug. 2020.

\bibitem{Intro_10}
R.~Fu, Y.~Tian, T.~Bao, Z.~Meng, and P.~Shi,
  ``\BIBforeignlanguage{en}{Improvement {Motor} {Imagery} {EEG}
  {Classification} {Based} on {Regularized} {Linear} {Discriminant}
  {Analysis}},'' \emph{\BIBforeignlanguage{en}{Journal of Medical Systems}},
  vol.~43, no.~6, p. 169, Jun. 2019.

\bibitem{Intro_11}
{Yinxia Liu}, {Weidong Zhou}, {Qi Yuan}, and {Shuangshuang Chen}, ``Automatic
  {Seizure} {Detection} {Using} {Wavelet} {Transform} and {SVM} in
  {Long}-{Term} {Intracranial} {EEG},'' \emph{IEEE Transactions on Neural
  Systems and Rehabilitation Engineering}, vol.~20, no.~6, pp. 749--755, Nov.
  2012.

\bibitem{Intro_12}
O.~W. Samuel, Y.~Geng, X.~Li, and G.~Li, ``\BIBforeignlanguage{en}{Towards
  {Efficient} {Decoding} of {Multiple} {Classes} of {Motor} {Imagery} {Limb}
  {Movements} {Based} on {EEG} {Spectral} and {Time} {Domain} {Descriptors}},''
  \emph{\BIBforeignlanguage{en}{Journal of Medical Systems}}, vol.~41, no.~12,
  p. 194, Dec. 2017.

\bibitem{Intro_13}
{Jie Xu}, {Yuan Yan Tang}, {Bin Zou}, {Zongben Xu}, {Luoqing Li}, and {Yang
  Lu}, ``The {Generalization} {Ability} of {Online} {SVM} {Classification}
  {Based} on {Markov} {Sampling},'' \emph{IEEE Transactions on Neural Networks
  and Learning Systems}, vol.~26, no.~3, pp. 628--639, Mar. 2015.

\bibitem{Intro_14}
C.-T. Lin, C.-H. Chuang, Y.-C. Hung, C.-N. Fang, D.~Wu, and Y.-K. Wang, ``A
  {Driving} {Performance} {Forecasting} {System} {Based} on {Brain} {Dynamic}
  {State} {Analysis} {Using} 4-{D} {Convolutional} {Neural} {Networks},''
  \emph{IEEE Transactions on Cybernetics}, pp. 1--9, 2020.

\bibitem{Intro_15}
S.~U. Amin, M.~Alsulaiman, G.~Muhammad, M.~A. Mekhtiche, and M.~Shamim~Hossain,
  ``\BIBforeignlanguage{en}{Deep {Learning} for {EEG} motor imagery
  classification based on multi-layer {CNNs} feature fusion},''
  \emph{\BIBforeignlanguage{en}{Future Generation Computer Systems}}, vol. 101,
  pp. 542--554, Dec. 2019.

\bibitem{Intro_16}
G.~Dai, J.~Zhou, J.~Huang, and N.~Wang, ``{HS}-{CNN}: a {CNN} with hybrid
  convolution scale for {EEG} motor imagery classification,'' \emph{Journal of
  Neural Engineering}, vol.~17, no.~1, p. 016025, Jan. 2020.

\bibitem{Intro_16.5}
\BIBentryALTinterwordspacing
Y.~Ding, N.~Robinson, Q.~Zeng, and C.~Guan, ``{TSception}: {Capturing}
  {Temporal} {Dynamics} and {Spatial} {Asymmetry} from {EEG} for {Emotion}
  {Recognition},'' \emph{arXiv:2104.02935 [cs]}, Apr. 2021, arXiv: 2104.02935.
  [Online]. Available: \url{http://arxiv.org/abs/2104.02935}
\BIBentrySTDinterwordspacing

\bibitem{Intro_17}
J.~He, L.~Zhao, H.~Yang, M.~Zhang, and W.~Li, ``{HSI}-{BERT}: {Hyperspectral}
  {Image} {Classification} {Using} the {Bidirectional} {Encoder}
  {Representation} {From} {Transformers},'' \emph{IEEE Transactions on
  Geoscience and Remote Sensing}, vol.~58, no.~1, pp. 165--178, Jan. 2020.

\bibitem{Intro_18}
T.~Zhang, W.~Zheng, Z.~Cui, Y.~Zong, and Y.~Li, ``Spatial–{Temporal}
  {Recurrent} {Neural} {Network} for {Emotion} {Recognition},'' \emph{IEEE
  Transactions on Cybernetics}, vol.~49, no.~3, pp. 839--847, Mar. 2019.

\bibitem{Intro_19}
N.~Zhang, ``Learning {Adversarial} {Transformer} for {Symbolic} {Music}
  {Generation},'' \emph{IEEE Transactions on Neural Networks and Learning
  Systems}, pp. 1--10, 2020.

\bibitem{Intro_20}
C.~S. Zandvoort, J.~H. van Dieën, N.~Dominici, and A.~Daffertshofer,
  ``\BIBforeignlanguage{en}{The human sensorimotor cortex fosters muscle
  synergies through cortico-synergy coherence},''
  \emph{\BIBforeignlanguage{en}{NeuroImage}}, vol. 199, pp. 30--37, Oct. 2019.

\bibitem{Intro_21}
\BIBentryALTinterwordspacing
A.~Vaswani, N.~Shazeer, N.~Parmar, J.~Uszkoreit, L.~Jones, A.~N. Gomez,
  L.~Kaiser, and I.~Polosukhin, ``Attention {Is} {All} {You} {Need},''
  \emph{arXiv:1706.03762 [cs]}, Dec. 2017, arXiv: 1706.03762. [Online].
  Available: \url{http://arxiv.org/abs/1706.03762}
\BIBentrySTDinterwordspacing

\bibitem{Intro_22}
S.~K. Goh, H.~A. Abbass, K.~C. Tan, A.~Al-Mamun, C.~Wang, and C.~Guan,
  ``Automatic {EEG} {Artifact} {Removal} {Techniques} by {Detecting}
  {Influential} {Independent} {Components},'' \emph{IEEE Transactions on
  Emerging Topics in Computational Intelligence}, vol.~1, no.~4, pp. 270--279,
  Aug. 2017.

\bibitem{Intro_23}
M.~Li and W.~Chen, ``\BIBforeignlanguage{en}{{FFT}-based deep feature learning
  method for {EEG} classification},'' \emph{\BIBforeignlanguage{en}{Biomedical
  Signal Processing and Control}}, vol.~66, p. 102492, Apr. 2021.

\bibitem{Intro_24}
P.~Kant, S.~H. Laskar, J.~Hazarika, and R.~Mahamune,
  ``\BIBforeignlanguage{en}{{CWT} {Based} {Transfer} {Learning} for {Motor}
  {Imagery} {Classification} for {Brain} computer {Interfaces}},''
  \emph{\BIBforeignlanguage{en}{Journal of Neuroscience Methods}}, vol. 345, p.
  108886, Nov. 2020.

\bibitem{Intro_25}
{Kai Keng Ang}, {Zhang Yang Chin}, {Haihong Zhang}, and {Cuntai Guan}, ``Filter
  {Bank} {Common} {Spatial} {Pattern} ({FBCSP}) in {Brain}-{Computer}
  {Interface},'' in \emph{2008 {IEEE} {International} {Joint} {Conference} on
  {Neural} {Networks} ({IEEE} {World} {Congress} on {Computational}
  {Intelligence})}.\hskip 1em plus 0.5em minus 0.4em\relax Hong Kong, China:
  IEEE, Jun. 2008, pp. 2390--2397.

\bibitem{Intro_26}
J.~Chen, Z.~Yu, Z.~Gu, and Y.~Li, ``Deep {Temporal}-{Spatial} {Feature}
  {Learning} for {Motor} {Imagery}-{Based} {Brain}–{Computer} {Interfaces},''
  \emph{IEEE Transactions on Neural Systems and Rehabilitation Engineering},
  vol.~28, no.~11, pp. 2356--2366, Nov. 2020.

\bibitem{Related_1}
F.~Lotte, M.~Congedo, A.~Lécuyer, F.~Lamarche, and B.~Arnaldi, ``A review of
  classification algorithms for {EEG}-based brain–computer interfaces,''
  \emph{Journal of Neural Engineering}, vol.~4, no.~2, pp. R1--R13, Jun. 2007.

\bibitem{Related_2}
F.~Lotte, L.~Bougrain, A.~Cichocki, M.~Clerc, M.~Congedo, A.~Rakotomamonjy, and
  F.~Yger, ``A review of classification algorithms for {EEG}-based
  brain–computer interfaces: a 10 year update,'' \emph{Journal of Neural
  Engineering}, vol.~15, no.~3, p. 031005, Jun. 2018.

\bibitem{Related_3}
S.~Sakhavi, C.~Guan, and S.~Yan, ``Learning {Temporal} {Information} for
  {Brain}-{Computer} {Interface} {Using} {Convolutional} {Neural} {Networks},''
  \emph{IEEE Transactions on Neural Networks and Learning Systems}, vol.~29,
  no.~11, pp. 5619--5629, Nov. 2018.

\bibitem{Related_4}
X.~Zhao, H.~Zhang, G.~Zhu, F.~You, S.~Kuang, and L.~Sun, ``A {Multi}-{Branch}
  {3D} {Convolutional} {Neural} {Network} for {EEG}-{Based} {Motor} {Imagery}
  {Classification},'' \emph{IEEE Transactions on Neural Systems and
  Rehabilitation Engineering}, vol.~27, no.~10, pp. 2164--2177, Oct. 2019.

\bibitem{Related_5}
L.~Yang, Y.~Song, K.~Ma, and L.~Xie, ``Motor {Imagery} {EEG} {Decoding}
  {Method} {Based} on a {Discriminative} {Feature} {Learning} {Strategy},''
  \emph{IEEE Transactions on Neural Systems and Rehabilitation Engineering},
  vol.~29, pp. 368--379, 2021.

\bibitem{Related_6}
P.~Wang, A.~Jiang, X.~Liu, J.~Shang, and L.~Zhang, ``{LSTM}-{Based} {EEG}
  {Classification} in {Motor} {Imagery} {Tasks},'' \emph{IEEE Transactions on
  Neural Systems and Rehabilitation Engineering}, vol.~26, no.~11, pp.
  2086--2095, Nov. 2018.

\bibitem{Related_7}
X.~Ma, S.~Qiu, C.~Du, J.~Xing, and H.~He, ``Improving {EEG}-{Based} {Motor}
  {Imagery} {Classification} via {Spatial} and {Temporal} {Recurrent} {Neural}
  {Networks},'' in \emph{2018 40th {Annual} {International} {Conference} of the
  {IEEE} {Engineering} in {Medicine} and {Biology} {Society} ({EMBC})}.\hskip
  1em plus 0.5em minus 0.4em\relax Honolulu, HI: IEEE, Jul. 2018, pp.
  1903--1906.

\bibitem{Related_8}
R.~Zhang, Q.~Zong, L.~Dou, X.~Zhao, Y.~Tang, and Z.~Li,
  ``\BIBforeignlanguage{en}{Hybrid deep neural network using transfer learning
  for {EEG} motor imagery decoding},'' \emph{\BIBforeignlanguage{en}{Biomedical
  Signal Processing and Control}}, vol.~63, p. 102144, Jan. 2021.

\bibitem{Related_9}
\BIBentryALTinterwordspacing
A.~Dosovitskiy, L.~Beyer, A.~Kolesnikov, D.~Weissenborn, X.~Zhai,
  T.~Unterthiner, M.~Dehghani, M.~Minderer, G.~Heigold, S.~Gelly, J.~Uszkoreit,
  and N.~Houlsby, ``An {Image} is {Worth} 16x16 {Words}: {Transformers} for
  {Image} {Recognition} at {Scale},'' \emph{arXiv:2010.11929 [cs]}, Oct. 2020,
  arXiv: 2010.11929. [Online]. Available: \url{http://arxiv.org/abs/2010.11929}
\BIBentrySTDinterwordspacing

\bibitem{Related_10}
X.~Zheng and W.~Chen, ``\BIBforeignlanguage{en}{An {Attention}-based
  {Bi}-{LSTM} {Method} for {Visual} {Object} {Classification} via {EEG}},''
  \emph{\BIBforeignlanguage{en}{Biomedical Signal Processing and Control}},
  vol.~63, p. 102174, Jan. 2021.

\bibitem{Related_11}
W.~Tao, C.~Li, R.~Song, J.~Cheng, Y.~Liu, F.~Wan, and X.~Chen, ``{EEG}-based
  {Emotion} {Recognition} via {Channel}-wise {Attention} and {Self}
  {Attention},'' \emph{IEEE Transactions on Affective Computing}, pp. 1--1,
  2020.

\bibitem{Related_12}
D.~Zhang, L.~Yao, K.~Chen, and J.~Monaghan, ``A {Convolutional} {Recurrent}
  {Attention} {Model} for {Subject}-{Independent} {EEG} {Signal} {Analysis},''
  \emph{IEEE Signal Processing Letters}, vol.~26, no.~5, pp. 715--719, May
  2019.

\bibitem{Related_13}
D.~Zhang, K.~Chen, D.~Jian, and L.~Yao, ``Motor {Imagery} {Classification} via
  {Temporal} {Attention} {Cues} of {Graph} {Embedded} {EEG} {Signals},''
  \emph{IEEE Journal of Biomedical and Health Informatics}, vol.~24, no.~9, pp.
  2570--2579, Sep. 2020.

\bibitem{Method_1}
J.~S. Kirar and R.~K. Agrawal, ``Relevant {Frequency} {Band} {Selection} using
  {Sequential} {Forward} {Feature} {Selection} for {Motor} {Imagery} {Brain}
  {Computer} {Interfaces},'' in \emph{2018 {IEEE} {Symposium} {Series} on
  {Computational} {Intelligence} ({SSCI})}.\hskip 1em plus 0.5em minus
  0.4em\relax Bangalore, India: IEEE, Nov. 2018, pp. 52--59.

\bibitem{Method_2}
\BIBentryALTinterwordspacing
K.~He, X.~Zhang, S.~Ren, and J.~Sun, ``Deep {Residual} {Learning} for {Image}
  {Recognition},'' \emph{arXiv:1512.03385 [cs]}, Dec. 2015, arXiv: 1512.03385.
  [Online]. Available: \url{http://arxiv.org/abs/1512.03385}
\BIBentrySTDinterwordspacing

\bibitem{Mehtod_3}
\BIBentryALTinterwordspacing
D.~Hendrycks and K.~Gimpel, ``Gaussian {Error} {Linear} {Units} ({GELUs}),''
  \emph{arXiv:1606.08415 [cs]}, Jul. 2020, arXiv: 1606.08415. [Online].
  Available: \url{http://arxiv.org/abs/1606.08415}
\BIBentrySTDinterwordspacing

\bibitem{Method_4}
\BIBentryALTinterwordspacing
J.~L. Ba, J.~R. Kiros, and G.~E. Hinton, ``Layer {Normalization},''
  \emph{arXiv:1607.06450 [cs, stat]}, Jul. 2016, arXiv: 1607.06450. [Online].
  Available: \url{http://arxiv.org/abs/1607.06450}
\BIBentrySTDinterwordspacing

\bibitem{Exp_1}
C.~Brunner, R.~Leeb, G.~R. Muller-Putz, and A.~Schlogl,
  ``\BIBforeignlanguage{en}{{BCI} {Competition} 2008 – {Graz} data set
  {A}},'' p.~6.

\bibitem{Exp_2}
R.~Leeb, C.~Brunner, G.~R. Muller-Putz, and A.~Schlogl,
  ``\BIBforeignlanguage{en}{{BCI} {Competition} 2008 – {Graz} data set
  {B}},'' p.~6.

\bibitem{Exp_3}
\BIBentryALTinterwordspacing
D.~P. Kingma and J.~Ba, ``Adam: {A} {Method} for {Stochastic} {Optimization},''
  \emph{arXiv:1412.6980 [cs]}, Jan. 2017, arXiv: 1412.6980. [Online].
  Available: \url{http://arxiv.org/abs/1412.6980}
\BIBentrySTDinterwordspacing

\bibitem{Exp_4}
K.~K. Ang, Z.~Y. Chin, C.~Wang, C.~Guan, and H.~Zhang, ``Filter {Bank} {Common}
  {Spatial} {Pattern} {Algorithm} on {BCI} {Competition} {IV} {Datasets} 2a and
  2b,'' \emph{Frontiers in Neuroscience}, vol.~6, 2012.

\bibitem{Exp_5}
R.~T. Schirrmeister, J.~T. Springenberg, L.~D.~J. Fiederer, M.~Glasstetter,
  K.~Eggensperger, M.~Tangermann, F.~Hutter, W.~Burgard, and T.~Ball,
  ``\BIBforeignlanguage{en}{Deep learning with convolutional neural networks
  for {EEG} decoding and visualization: {Convolutional} {Neural} {Networks} in
  {EEG} {Analysis}},'' \emph{\BIBforeignlanguage{en}{Human Brain Mapping}},
  vol.~38, no.~11, pp. 5391--5420, Nov. 2017.

\bibitem{Exp_6}
V.~J. Lawhern, A.~J. Solon, N.~R. Waytowich, S.~M. Gordon, C.~P. Hung, and
  B.~J. Lance, ``{EEGNet}: a compact convolutional neural network for
  {EEG}-based brain–computer interfaces,'' \emph{Journal of Neural
  Engineering}, vol.~15, no.~5, p. 056013, Oct. 2018.

\bibitem{Exp_7}
R.~Zhang, Q.~Zong, L.~Dou, and X.~Zhao, ``A novel hybrid deep learning scheme
  for four-class motor imagery classification,'' \emph{Journal of Neural
  Engineering}, vol.~16, no.~6, p. 066004, Oct. 2019.

\bibitem{Exp_8}
X.~Tang, W.~Li, X.~Li, W.~Ma, and X.~Dang, ``\BIBforeignlanguage{en}{Motor
  imagery {EEG} recognition based on conditional optimization empirical mode
  decomposition and multi-scale convolutional neural network},''
  \emph{\BIBforeignlanguage{en}{Expert Systems with Applications}}, vol. 149,
  p. 113285, Jul. 2020.

\bibitem{Exp_9}
L.~van~der Maaten and G.~Hinton, ``Visualizing data using t-sne,''
  \emph{Journal of Machine Learning Research}, vol.~9, no.~86, pp. 2579--2605,
  2008.

\bibitem{Exp_10}
J.~Hu, L.~Shen, and G.~Sun, ``Squeeze-and-{Excitation} {Networks},'' in
  \emph{2018 {IEEE}/{CVF} {Conference} on {Computer} {Vision} and {Pattern}
  {Recognition}}.\hskip 1em plus 0.5em minus 0.4em\relax Salt Lake City, UT:
  IEEE, Jun. 2018, pp. 7132--7141.

\bibitem{Exp_11}
\BIBentryALTinterwordspacing
M.~Lin, Q.~Chen, and S.~Yan, ``Network {In} {Network},'' \emph{arXiv:1312.4400
  [cs]}, Mar. 2014, arXiv: 1312.4400. [Online]. Available:
  \url{http://arxiv.org/abs/1312.4400}
\BIBentrySTDinterwordspacing

\end{thebibliography}
\end{document}